\documentclass[preprint]{interact}

\usepackage{units,amsmath,amssymb}
\usepackage{subfigure}
\usepackage{graphicx,xcolor}

\usepackage{eurosym}

\usepackage{float}

\graphicspath{{Kuvat/}}

\usepackage[utf8]{inputenc}
\usepackage[T1]{fontenc}
\usepackage{textcomp}
\usepackage{gensymb}

\usepackage[round]{natbib}

\newcommand{\be}{\begin{equation}}
\newcommand{\ee}{\end{equation}}

\newcommand{\bea}{\begin{eqnarray}}
\newcommand{\eea}{\end{eqnarray}}

\usepackage[normalem]{ulem} 
\newcommand\hl{\bgroup\markoverwith
	{\textcolor{yellow}{\rule[-.5ex]{2pt}{2.5ex}}}\ULon}

\begin{document}

\title{
Energy analysis in ice hockey arenas and analytical formula for the temperature profile
in the ice pad with transient boundary conditions
}

\author{
	\name{Andrea Ferrantelli\textsuperscript{a}\thanks{CONTACT Andrea Ferrantelli. Email: andrea.ferrantelli@taltech.ee}, Klaus Viljanen\textsuperscript{b} and Jarek Kurnitski\textsuperscript{a,b}}
	\affil{\textsuperscript{a}Tallinn University of Technology, Department of Civil Engineering and Architecture, 19086 Tallinn, Estonia; \textsuperscript{b}Aalto University, Department of Civil Engineering, P.O.Box 12100, 00076 Aalto, Finland}
}

\maketitle

\date{\today}

\begin{abstract}
	
The energy efficiency of ice hockey arenas is a central concern for the administrations, as these buildings are well known to consume a large amount of energy. Since they are composite, complex systems, solutions to such a problem can be approached from many different areas, from managerial to technological to more strictly physical.

In this paper we consider heat transfer processes in an ice hockey hall, during operating conditions, with a bottom-up approach based upon on-site measurements. Detailed heat flux, relative humidity and temperature data for the ice pad and the indoor air are used for a heat balance calculation in the steady-state regime, which quantifies the impact of each single heat source. We then solve the heat conduction equation for the ice pad in transient regime, and obtain a generic analytical formula for the temperature profile that can be used in practical applications.

We then apply this formula to the resurfacing process for validation, and find good agreement with an analogous numerical solution. Since it is given with implicit initial condition and boundary conditions, it can be used not only in ice hockey halls, but in a large variety of engineering applications.

\end{abstract}




\section{Introduction}

A well-known major preoccupation of contemporary building research is the energy efficiency of buildings, which is subject to increasingly stringent requirements set by manufacturers and international institutions (\cite{EU,doi:10.1080/17512549.2012.756430}) under ecological and economical motivations.

In this context, ice hockey arenas are fairly demanding systems, infamous for consuming a very large amount of energy that approaches $\sim 1800$ MWh per year. In particular, according to several studies (see e.g. \cite{Karampour,Miska,opt1}) refrigeration is responsible for nearly 43 percent of the total energy consumption of the ice hockey hall. Process optimization and energy efficiency have therefore increased their role of key research concepts on very diverse, yet correlated aspects of energy management.
	These can range from the concrete pad design as in \cite{Haghighi}, to the air distribution and ventilation (\cite{Yang,doi:10.1034/j.1600-0668.2001.110206.x}) and their effect on the overall heat balance, as addressed by \cite{PALMOWSKA2018373,toomla2018experimental,en12040693}, as well as the exergoeconomic analysis performed by \cite{EROL2017118}.

		Furthermore, these types of buildings constitute a challenging setup for studies in building physics, due to the diverse physical processes often concurring to each other, which take place in the environment. For instance, ventilation and air conditioning in such large and complex systems are an intriguing field of study, as illustrated by \cite{Caliskan20101418,Daoud20081782,Teyssedou2013229}. Moreover, opportunities for energy conversion (\cite{Zhao2016}), electrical energy storage (\cite{chen2013compressed}) and passive measures such as air transport and water vapour control (\cite{doi:10.1080/17512549.2018.1488617}) find in this context an interesting field of study and application.
	All in all, very different types of solutions can be realized for promoting the energy saving by increasing the overall system efficiency, see e.g. \cite{Castellani_2017} for solar energy storage and \cite{Castellani_2019} for pure methane collection.
	
	The various contributions to the energy balance in such a complex system are usually addressed numerically.
	Heat and mass transfer processes occur both above and inside the large ice/concrete slab forming the ice rink, and have been thoroughly studied for at least a couple of decades now, see for instance \cite{Negiz,Hastaoglu} and more recently \cite{Teyssedou}.
	Computational Fluid Dynamics (CFD), see e.g. \cite{doi:10.3763/aber.2009.0405} for a review, is a useful tool in this sense. CFD simulations in 2D were performed under steady state conditions in \cite{Bellache2005417}, where velocity, temperature and absolute humidity distributions were predicted for an indoor ice rink with heating provided by ventilation. The numerical model was then extended to transient processes by the same authors in \cite{2288256320061001}, adding calculations of heat transfer through the ground, energy gains from lights and resurfacing effects.	
	A CFD numerical analysis of the radiative component induced by thermostatically controlled radiant heaters has been considered in \cite{Omri2016}, tracking the heat flow and temperature inside the ice hall, together with the heat fluxes into the ice pad.

	Regarding analytical solutions, the literature is far less abundant mostly because of evident technical difficulties. General solutions for coupled heat transfer do exist however, since they pertain more general cases such as building foundations and radiant floors.
	\cite{Somrani20081687} used the Interzone Temperature Profile Estimation (ITPE) method developed in \cite{krarti1999building}, considering an ice pad with constant upper boundary condition (the air temperature just above the ice surface), laying over sand, insulation, a soil layer with time-varying temperature and a water table with constant temperature.
	
More recent investigations concentrate on large scales, and consider the entire ice hall in view of numerical integration towards energy optimization, see for instance \cite{MUN20111087}.
	 As an example, \cite{doi:10.1177/0144598717723644} formulated a model for assessing the thermal performance of a cooling system consisting of an ice rink, a chiller unit and a spherical thermal energy storage tank. Computing the ice rink heat gain and the energy consumption of the chiller unit, they were able to determine the operation time of this large system in function of indoor conditions, storage tank and chiller characteristics.

Unfortunately, heat flux data during the ice pad resurfacing, which constitutes the more energy consuming event in the entire operative cycle, are still missing from the literature. Additionally, in this specific context there currently exists a lack of heat transfer formulas which are easy to apply to practical design problems.
	
	In this paper we try to address these issues by considering the topic on a smaller scale, i.e. an ice pad element, by means of a bottom-up approach based on measurements. For the first time in the literature, we report detailed measurements of the heat flux as well as ice temperature at the surface and at the bottom of the ice pad during resurfacing. These constitute the time-dependent boundary conditions (b.c.) for an analytical temperature profile along the ice slab depth which we accordingly compute.

	Though the derivation is rather involved, we recast the resulting temperature profile in function of the boundary conditions as a simple formula which can be easily implemented in common computational software. This is another novelty of this paper, aimed at immediate practical usage.
 Our main purpose is indeed to provide a concise reference for practical applications in ice hockey halls, by means of an easily applicable formula for the ice pad temperature together with comprehensive data of heat flux, air stratification and humidity at various heights in the ice hockey arena.

	We choose the resurfacing stages because they are the most complex and energy consuming phases of the operational cycle of an ice hockey hall, due to the large amount of heat transferred to the ice pad in a relatively short time. This affects importantly the efficiency and energy consumption of the refrigerating system, and here we aim at finding applicable quantitative knowledge to aid e.g. system control and space heating investigations.

	Though obtained for a specific case-study, our measurements can be applied to the majority of standard ice hockey halls, for instance in the development of new control methods for ice rink cooling systems (see for instance \cite{Lu201491,Lu2015261}). Our findings also help in developing methods to reduce the indoor temperature stratification as suggested by \cite{en12040693}, or to collect a portion of the heat generated and reuse it in the ice hockey hall. Moreover, the temperature profile formula we obtain is very general and is not limited to ice hockey arenas; rather, it can be applied to other heat conduction processes with transient boundary conditions.

	The present paper is organized as follows:
	Section \ref{measurements} reports the experimental setup and measurements, as well as an energy balance analysis which is validated with experimental data. We examine the heat flux on the ice pad, the indoor air relative humidity (RH) in its proximity and the temperatures at surface and at the ice/concrete interface during a typical day of operation.

	In Section \ref{steadystate} we use the field measurements to estimate the diverse contributions to the heat load over the ice pad in the steady state conditions immediately preceding the resurfacing, considering convection, condensation and	radiation. 
	An analytical formula for the ice pad temperature in transient conditions is then derived in Section \ref{analytical}. Our results are summarized and further discussed in the Conclusions, and finally in Appendix A we apply our formula to the calculation of the temperature profile inside the ice pad during resurfacing, validating it with the initial data.

\begin{figure}[t]
\centering
\includegraphics[width=0.7\textwidth]{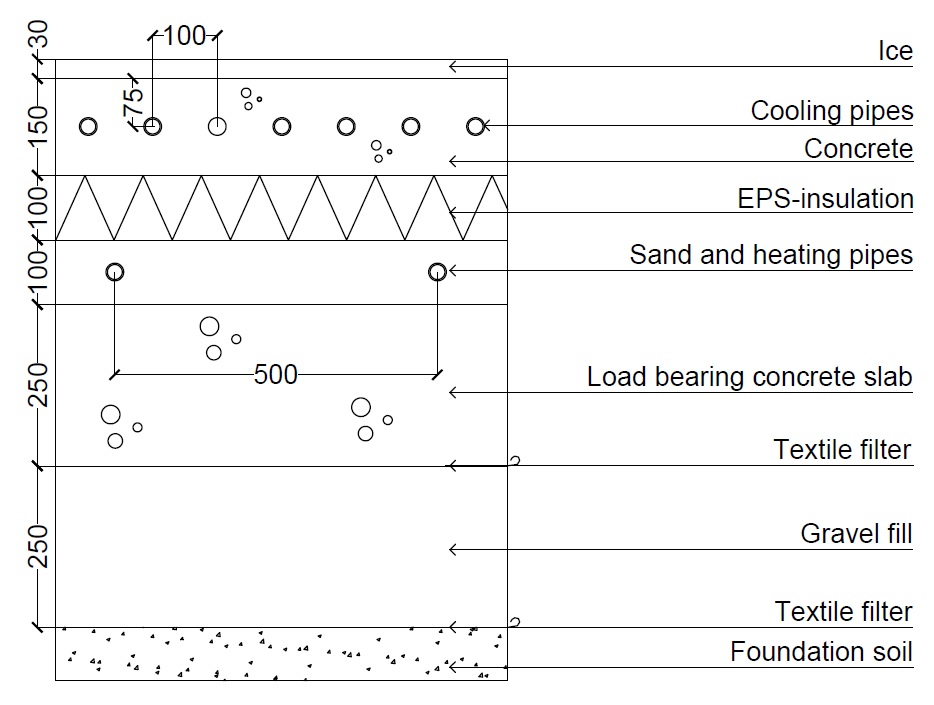}   
\caption{A portion of the ice rink floor structure considered.}
\label{fig:slab}

\end{figure}

\section{Ice hockey hall and temperature measurements}\label{measurements}
%
%
%

	Good skating conditions occur when the ice surface is smooth, hard and slippery enough. Usually the ice temperature at the surface is between -3\textdegree C and -5\textdegree C. In any cases the temperature of the ice should be kept under -1\textdegree C to maintain ice hardness, as explained in \cite{ICEH}. To keep the ice smooth and in optimal conditions after the wear due to skating, it is necessary to perform periodical maintenance, which is usually done by means of resurfacing machines. These devices first shave the ice, then brush it and eventually spread a thin layer of new water on the surface.
	
	At each maintenance cycle, 300 to 800 liters of water are used, corresponding to a 0.25-0.5mm thick water layer. The water cooling and freezing processes generate a sudden increase in the ice temperature, which lowers the cooling efficiency of the refrigerating system, thus increasing the operational costs of the ice hall. Accurate analysis of this phenomenon is then important to optimize the refrigeration process and to achieve overall energy saving.
	
	The field study discussed in this paper was carried out at the Reebok Arena in Lepp\"avaara, Finland, during summer conditions\footnote{\cite{toomla2018experimental} have performed an extended study in a different ice hockey arena.}. The ice hall has two identical ice rinks of size 1624 m2 each. There are no major stands in the hall and the ventilation air flow rate varies between 3-6 m3/s during the year. The refrigeration system is indirect, with ammonium as the refrigerant and ethylene glycol as the brine.
	
	Consider now a modular component of the ice/concrete slab as in Fig.\ref{fig:slab}: ice 30mm, concrete slab 150mm with brine pipes at the middle, EPS-insulation 100mm, sand 100mm with frost pipes and a load bearing concrete slab 250mm. The coefficient of performance for the refrigeration system is $\sim1.6$ on the average, as documented in \cite{opt1}.

\begin{figure}[t]
\centering
\includegraphics[width=0.8\textwidth]{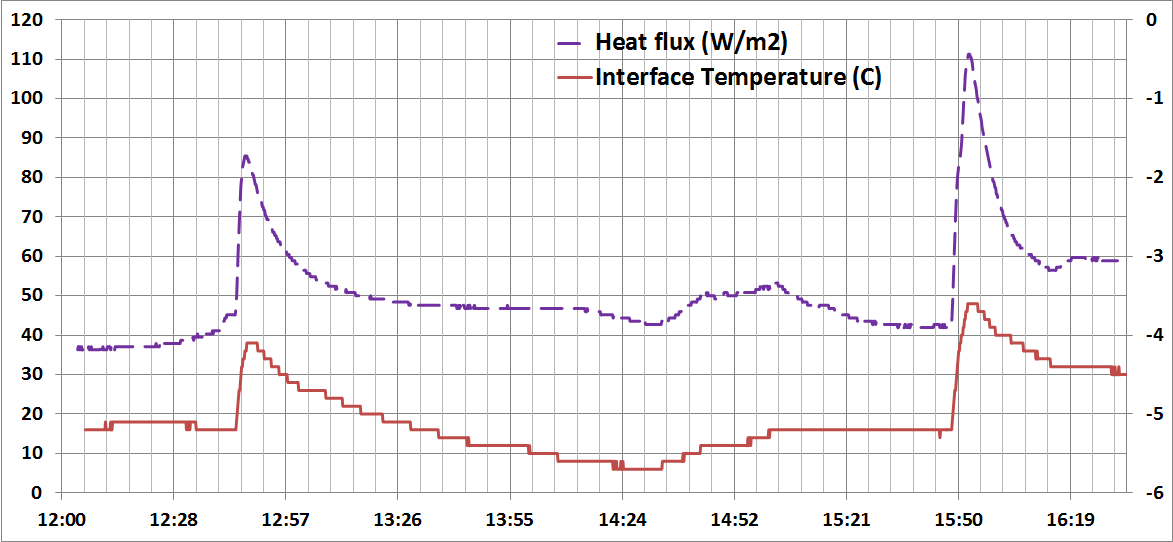}
\caption{Measurements of temperature and heat flux at the ice/concrete interface.}
\label{fig:Tqgeneral}
\end{figure}

\begin{figure}[t]
		\centering
		\includegraphics[width=0.8\textwidth]{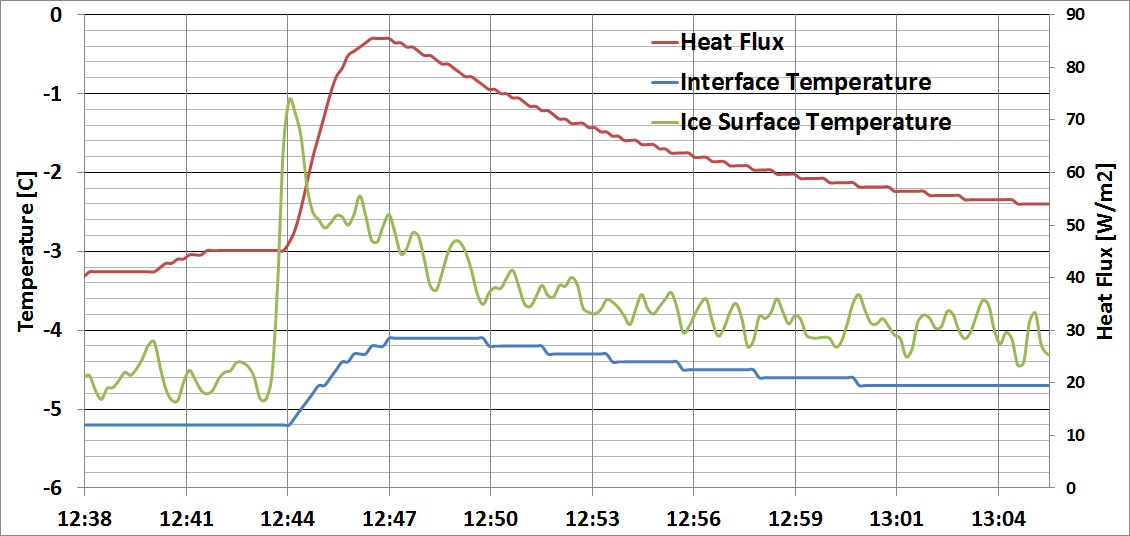} 
		\caption{Heat flux and temperatures during the water chilling and freezing.}
		\label{fig:Tqzoom}
	\end{figure}

	We focus on the first resurfacing process in Fig.\ref{fig:Tqgeneral}, corresponding to the peak near 12:45. The ice surface has an initial surface temperature $T_S\sim$-4.5\textdegree C, the whole ice pad is at $-5$\textdegree C on the average.
	During resurfacing, $m_w$=450kg of water at temperature $T_ w$=40\textdegree C are spread on the ice surface at $t$=0s. The entire process consists of three phases:

1. cooling of the water layer by contact with the ice pad, convection and radiation with the surroundings,

2. complete freezing of the water by the same processes,

3. cooling of the new ice from $T_{freeze}$=0\textdegree C to $T_{ice}\sim$-4\textdegree C (see Fig.\ref{fig:Tqzoom}).
A thermal camera records the surface temperature point wise, taking pictures at intervals $\Delta t$=10s. Moreover, heat flux and temperature are measured with a heat flux plate and a pt-100 temperature sensor installed at the ice/concrete interface in the slab, as shown in Fig.\ref{fig:Tqgeneral}.
The according data are plotted in Fig.\ref{fig:Tqzoom}.

	The physical properties of water are evaluated at the film temperature $T_{f}$=20\textdegree C, implying a specific heat $c_{p,w}$=4.182 kJ/kgK.
%
Since $m_w$=450kg of water are spread on the ice, the average thickness is only
\be
x=\frac{m_w}{\rho A}=0.28\,mm\,,
\ee
which is anyway just an indicative value, as the precision of the resurfacing machine is not high enough.
	The size of the ice rink is $A$=1624 m2, $h_{fs}$=338 kJ/kg is the water latent heat of freezing and $c_{p,i}$=2.05 kJ/kgK is the specific heat of ice at $T\sim$0\textdegree C. The water chilling and freezing loads are thus obtained as follows,
	\bea
	&&Q_w=[m_w[h_{fs}+c_{p,w}(T_w-0)+c_{p,i}(0-T_{ice})]]
	\label{totalenergy}\\
	&&=231.21\,MJ\,;
	\; q_w=142.37\frac{kJ}{m^2}\,,
	\label{freezingenergy}\\
	&&Q_1= m_wc_{p,w}(T_w-0)=75.28\,MJ\,;
	\qquad
q_1=\frac{Q_1}{A}=46.4\,\frac{kJ}{m^2}\,,
	\label{Q1}\\
	&&Q_2= h_{fs}m_w=152.1\,MJ\,;
	\qquad
	q_2=\frac{Q_2}{A}=93.7\,\frac{kJ}{m^2}\,,
	\label{Q2}\\
	&&Q_3=m_wc_{p,i}(0-T_{ice}) =3.69\,MJ\,;
	\qquad
	q_3=\frac{Q_3}{A}=2.27\,\frac{kJ}{m^2}\,,\label{Q3}
	\eea
where $T_{ice}$ is the temperature of the frozen resurfacing water under the new steady state conditions.
The theoretical heat load Eq.(\ref{freezingenergy}) is consistent with the literature, as in \cite{Seghouani2009500,Seghouani2011383}.
%
%
%

\begin{figure}
\centering
\begin{minipage}[b]{0.45\linewidth}
\centering
		\includegraphics[width=\textwidth]{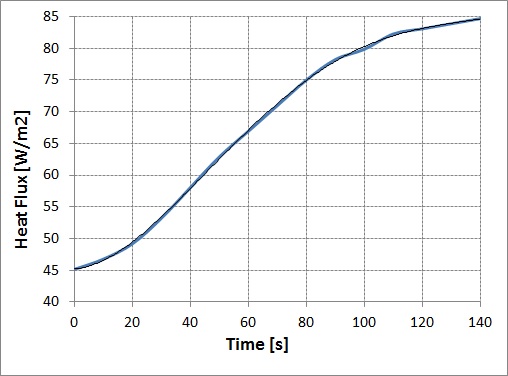}
		\caption{Interpolation curve for Eq.(\ref{Aint}).}
		\label{fig:HeatFlux_part1}
	\end{minipage}
	\hspace{0.2cm}
\begin{minipage}[b]{0.45\linewidth}
		\centering
		\includegraphics[width=\textwidth]{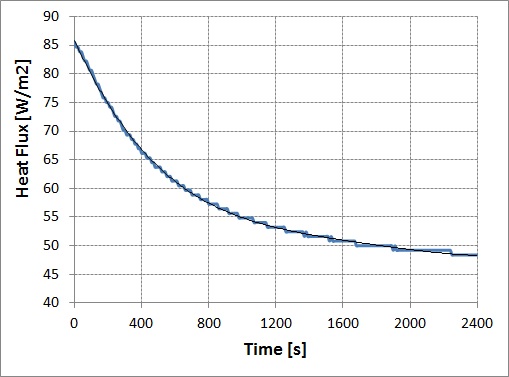}
		\caption{Interpolation curve for Eq.(\ref{Bint}).}
		\label{fig:HeatFlux_part2}
	\end{minipage}
\end{figure}

	Let us verify immediately that the theoretical heat flux in (\ref{freezingenergy}) is consistent with the measurements at the ice/concrete interface in Fig.\ref{fig:Tqzoom}. The integral of the heat flux curve 
is the total heat transferred during the resurfacing, to be compared with Eq.(\ref{freezingenergy}). For better precision, we split the curve into three contributions: two polynomials\footnote{At first sight the plot in Fig.\ref{fig:HeatFlux_part2} seems to suggest an exponential, however the accuracy is that case would be much lower, with $R^2=0.8552$.} plotted in Fig.\ref{fig:HeatFlux_part1} ($R^2=0.9997$) and Fig.\ref{fig:HeatFlux_part2} ($R^2=0.9991$), and a constant value $\dot{q}_{max}$=85.4839 W/m2, corresponding to the narrow ($\Delta t \sim$ 30s) plateau at the top of the curve in Fig.\ref{fig:Tqzoom}.

	If $\dot{q}_A$ is the heat flux rate for the first contribution and $\dot{q}_B$ the flux for the second curve, we obtain the following expressions:
\bea
\dot{q}_A(t)&=&3\times 10^{-7}t^4- 10^{-4}t^3+0.0106t^2+0.0414t+45.191\,,\\
\dot{q}_B(t)&=&3\times 10^{-12}t^4-2\times 10^{-8}t^3+5\times 10^{-5}t^2-0.0647t+85.873\,.
\eea
Integrating the above over the respective time intervals gives
\bea
&&q_A=\int^{140}_{0} \dot{q}_A(t)dt=10.05\,\frac{kJ}{m^2}\,,\label{Aint}\\
&&q_B=\int^{2300}_{0} \dot{q}_B(t)dt=127.88\,\frac{kJ}{m^2}\,,\label{Bint}
\eea
to which we add the heat transferred at the peak, namely
\be
q_{max}=85.4839\,\frac{W}{m^2}\times \Delta t\,[s]=2.57\,\frac{kJ}{m^2}\,.
\ee
Consistently with point 3. of the resurfacing process, the upper limit of the integral (\ref{Bint}) coincides with the instant when the ice surface temperature reaches $T_{ice}$=-4\textdegree C.

The total heat transferred to the ice pad during the three phases of resurfacing in Eq.(\ref{freezingenergy}) is therefore estimated as
\be
q_{exp}=140.49\,\frac{kJ}{m^2}\,.
\ee
This is very close to the theoretical mean value in Eq.(\ref{freezingenergy}), holding as
\be
q_w=142.37\,\frac{kJ}{m^2}\,,
\ee
we find indeed
\be
\Delta q\equiv q_w-q_{exp}=1.88\,\frac{kJ}{m^2},\qquad \%(\Delta q)=1.32\%\,,
\ee
which is fairly satisfactory and verifies the consistency of the theoretical and measured heat load\footnote{This is precise enough for our purposes, notice anyway that one has some freedom in setting the upper limit of integration in (\ref{Bint}). The order of magnitude is anyway what matters.}.

\subsection{Indoor air measurements}\label{sec:indoor-air-measurements}

Regarding the air temperature and RH, we used different sets of data obtained in different sessions.
One set of data was obtained at 0.04m, 0.1m and 0.24m above the ice rink with accuracy $\pm2\%$ for 0-90$\%$ relative humidity and $\pm3\%$ for 90-100$\%$ RH. The accuracy of the temperature measurement was $\pm$0.4\textdegree C.
Later on, additional measurements were made at 0.6m, 1.2m, 1.8m and 2.2m above the ice with PT-100 temperature sensors, with accuracy $\pm$0.5\textdegree C.
	Moreover, the temperature was measured at 5.0m and 8.3m height with an accuracy of $\pm$0.5\textdegree C. All measurements were done with $\Delta t$=10s interval and lasted approximately four hours. The probes set at 0.04-2.2m height were set up in a metal holder standing above the ice; this was located at one corner of the ice rink, 1m from the edge of the ice pad.

Further measurements were later performed at 0.005m, 0.01m, 0.04m, 0.1m, 0.24m and 5m height over the ice with different probes, to substantially reduce the errors. These now held as $\pm$0.3\% for RH and $\pm$0.1\textdegree C for the temperature, with an according $\pm$0.02 g/m3 error for the computed absolute humidity.
	While during resurfacing and skaters' practice such high accuracy is not necessarily needed, as the stratification is erased by the air mixing, under steady state conditions this is crucial for the energy balance estimate we perform in Section \ref{steadystate}. These later results are listed in Table \ref{table:RH}.

All data were collected at one instance after the sensors had became steady. The probes at 0.005-0.24m height were again set up in a metal holder over the ice. Two locations of the holder were used, at the corner of the ice rink and closer to the centre of the ice pad, at approximately 8m from the edge where also the heat flux plates were installed and thermal camera measurements were taken.

\begin{figure}[ht]
\centering
\includegraphics[width=0.8\textwidth]{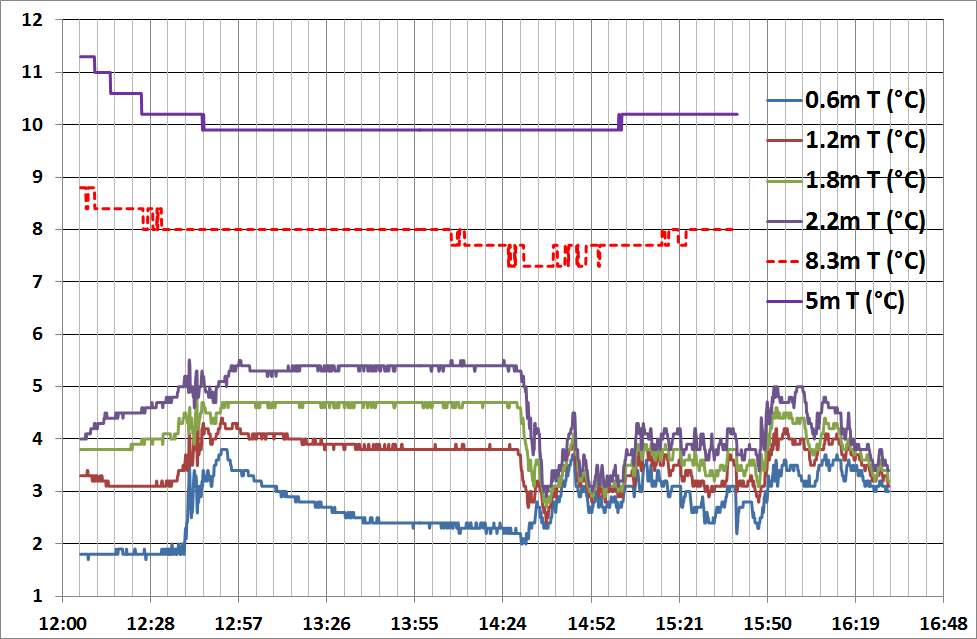}   
\caption{Air temperature above the ice during resurfacing and skaters activity.}
\label{fig:AirT}
\vspace{0.3cm}
\includegraphics[width=0.8\textwidth]{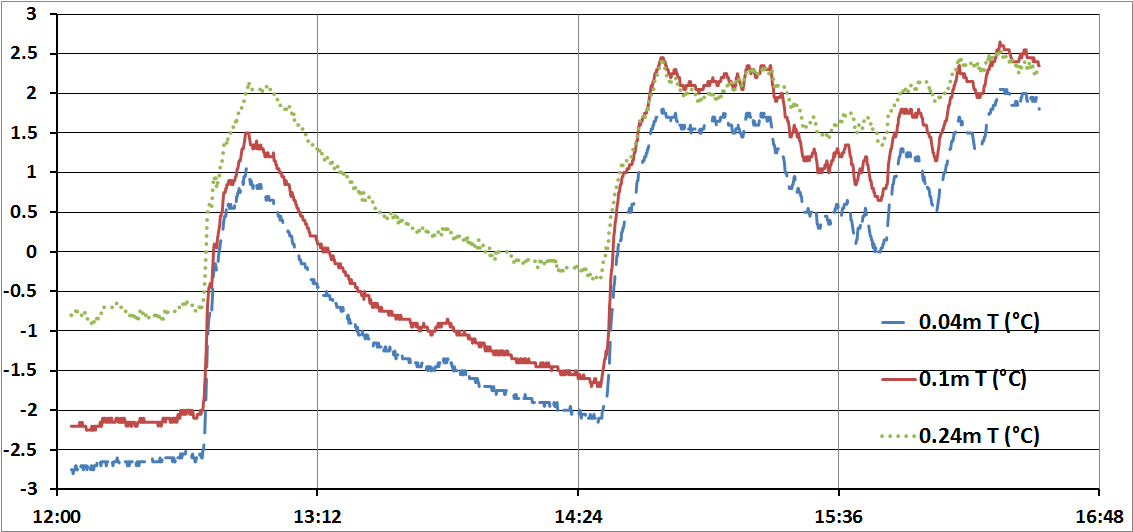}   
\caption{Air temperature above the ice rink at different heights, closer to the ice.}
\label{fig:AirTzoom}
\end{figure}

The air temperature at 0.6m-8.3m over the ice is plotted in Fig.\ref{fig:AirT}, while that at 0.04m-0.24m in Fig.\ref{fig:AirTzoom}. All the other data are shown in Table \ref{table:RH}. 

	The plots show that, as expected, the air temperature above the ice rink is strongly stratified, see Fig.\ref{fig:AirT}, Fig.\ref{fig:AirTzoom} and Table \ref{table:RH}, consistently with \cite{toomla2018experimental}. The temperature at 0.6m height is 2\textdegree C and 10\textdegree C at 5m. Interestingly, by virtue of the mixing effect of the ventilation system supplying warm air at 25\textdegree C at 5m above the ice, the highest temperature is at 5m instead of 8.3m (Fig.\ref{fig:AirT}). Furthermore, the stratified air temperature at 0.6-2.2m turns into a more uniform temperature when the skaters enter the ice rink at 14:30 and start mixing the air layers.

	The measured relative humidity (RH) and absolute humidity (AH), listed in Table \ref{table:RH}, suggest that above the ice rinks the RH is uniform at heights 0.005m - 0.24m over the ice and distinctly lower farther from the ice pad. The relative humidity is higher above the ice than in the other parts of the space, yet it never reaches 95-100\%.

The measurements did not imply any significant differences in temperature or relative humidity depending on the location over the ice pad, either corner or centre, in agreement with \cite{en12040693}.

\begin {table}[H]
\centering
\caption{Air temperature, relative humidity (RH) and absolute humidity (AH) as measured at different heights from the ice pad surface during steady state conditions.}
\begin{tabular}{ lccc  }
  \toprule
 Height over the ice [$\pm$0.005 m] & T [$\pm$0.1 \textdegree C] & RH [$\pm$0.3\%] & AH [$\pm$0.02 g/m3]\\
 \midrule
 5.0   & 4.2    &77.2  &   4.98 \\
 0.24&   -0.8  & 90.2  &4.15 \\
 0.1 &-2.2 & 89.6  &  3.76 \\
 0.04    &-2.7 & 89.3  &  3.59 \\
 0.01&   -3.1  & 88.1  &3.45 \\
 0.005& -3.5  & 88.2  &3.36 \\
\bottomrule
\end{tabular}

\label{table:RH}
\end{table}


\section{Energy balance and ice temperatures before resurfacing}\label{steadystate}

In this section we use the field measurements to compute some estimates of the heat loads on the ice pad before resurfacing. We consider convection, condensation and radiation. In this case the heat transfer is steady state in good approximation, so this is easy to accomplish. Even though it is difficult to obtain very precise values, this shows well the various factors concurring to the overall energy balance on the ice hockey rink.

	Consider first Fig.\ref{fig:Tqgeneral} and Fig.\ref{fig:Tqzoom} before resurfacing, namely before $\sim$12:43. The heat flux through the ice pad and the ice temperature, both measured at the ice/concrete interface, are respectively $\dot{q}$=41.85 W/m2 (average) and $T_I$=-5.2\textdegree C. The ice pad thickness is on the average $L$=30mm.
	We can derive the ice temperature at surface very easily, if $k_{ice}$=2.25 W/mK,
\be\label{icesurf}
T_S=T_I+\frac{L}{k_{ice}}\dot{q}=-4.64\,\mathrm{^{\circ} C}\,,
\ee
which is consistent with the ice surface temperature given in Fig.\ref{fig:Tqzoom}.

	The heat flux on the ice track surface is the sum of several contributions,
\bea\label{qsurf}
&&\dot{q} = \dot{q}_{rad}+\dot{q}_{conv}+\dot{q}_{wvcond}+\dot{q}_{lamp}\nonumber\\
&& = h_{rad}(T_{ceiling}-T_S)+(h_{conv}+h_{wvcond})(T_{in}-T_S)+\dot{q}_{lamp}\,,
\eea
namely thermal radiation from the ceiling, convection and water vapor condensation at the surface, and heat load from the lighting system. For simplicity, we neglect the thermal radiation from the vertical walls and from the audience stands. These should not give a relevant contribution, since each ice rink is much larger than the walls' height, and ice pad and walls are separated by several meters of empty floor. Additionally, the ice surface temperature was measured at $\sim$8m from the edge towards the ice pad's centre (Section \ref{sec:indoor-air-measurements}), thus farther away from the walls (see also \cite{Karampour}). This is fairly different from the situation addressed by \cite{Omri2016}, in a much smaller ice hockey hall.

	For convection we use $T_{in}=T$(0.005m)= -3.5\textdegree C, see Table \ref{table:RH}. Thus the heat transfer coefficient takes into account both natural convection and a correction given by forced convection, following \cite{ASHRAE10}
	\be
	h_{conv}=3.41+3.55\,V=3.94\frac{W}{m^2K}\,,
	\ee
	corresponding to $V=0.15\,m/s$ for the air flow right on top of the ice \cite{Miska}. The convection heat flow is therefore
%
\be\label{qconv}
\dot{q}_{conv}=h_{conv}(T_{in}-T_S)=4.49\frac{W}{m^2}\,.
\ee
	The radiation heat transfer coefficient is written instead as
\be\label{radcoeff}
h_{rad}= \varepsilon_{12}\,\sigma\,(T^2_{ceiling}+T^2_{ice.sur})(T_{ceiling}+T_S)=1.39\,\frac{W}{m^2K}\,,
\ee
where $T_{ceiling}$=18\textdegree C and the resulting emissivity is computed according to \cite{ASHRAE94},
\be\label{emiss}
\varepsilon_{12}=\left[\frac{1}{F_{ci}}+\left(\frac{1}{\varepsilon_{ceiling}}-1\right)+\frac{A_c}{A_i}\left(\frac{1}{\varepsilon_{ice.sur}}-1\right)\right]^{-1}\,.
\ee
The view factor between ceiling and ice is $F_{ci}=0.68$, while emissivities $\varepsilon_{ice.sur}=0.98$ and $\varepsilon_{ceiling}=0.33$ for ice surface and ceiling (a load bearing sheet of galvanized steel whose portion above the ice rink has area $A_c$=2600 m2) are respectively used, following \cite{Miska}. These give $\varepsilon_{12}=0.278$.

\begin{figure}[t]
		\centering
		\includegraphics[width=0.8\textwidth]{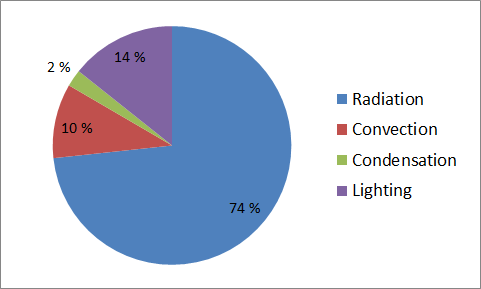} 
		\caption{Contributions to the heat load over the ice pad in steady state conditions, Eq(\ref{heatbalance}).}
		\label{fig:pie}
	\end{figure}

	One must also take into account the radiative heat transferred to the ice pad by the lighting system. The lamps in Lepp\"avaara are metal halide, which implies a contribution of 400W per lamp. The portion of this power that is turned into heat is nearly 62\%, as given by the manufacturer \cite{Osram}. Using the upper limit for the heat generation for 40 lamps gives the following contribution:
	\be
	\dot{Q}_{lamp}=9.92\,kW\,,   
	\ee
corresponding to the following heat flux,
\be\label{lamp}
\dot{q}_{lamp}=\frac{\dot{Q}_{lamp}}{A}=6.11\,\frac{W}{m^2}\,.
\ee
	The water vapor condensation heat load is computed following \cite{Granryd},
	\be
	\dot{q}_{wvcond}=h_d(T_{in}-T_S)\qquad \left[ \frac{W}{m^2}\right]\,,
	\ee
where the heat transfer coefficient for condensation $h_d$ is calculated from
\bea
h_d&=&1750 h_{conv} \frac{\Delta p}{\Delta T}\,,\qquad [p]=[atm] \nonumber\\
\Delta p &=& \varphi_{in}p_{in}-p_s\,.
\eea
Here $\Delta T=T_{in}-T_S$, and $\varphi_{in}= 0.88$ is the relative humidity at 5mm from the ice surface, as in Table \ref{table:RH}. The saturation pressures are calculated from (here [T]=[\textdegree C])
\bea
&& p_{in}=10^5 \exp\left( 17.391-\dfrac{6142.83}{273.15+T_{in}} \right)\,,\\
&& p_s=10^5 \exp\left( 17.391-\dfrac{6142.83}{273.15+T_S} \right)\,.
\eea
We thus obtain $h_d=0.9\,W/m^2K$ and $\dot{q}_{wvcond}=1.03\, W/m^2$.
	By substituting this result into (\ref{qsurf}), together with Eqs.(\ref{qconv}), (\ref{radcoeff}) and (\ref{emiss}), and adding also (\ref{lamp}), we get
	\be\label{heatbalance}
	\dot{q}=\dot{q}_{rad}+\dot{q}_{conv}+\dot{q}_{wvcond}+\dot{q}_{lamp}=31.25+4.49+1.03+6.11=42.87\frac{W}{m^2}\,,
	\ee
that overestimates only slightly the measured value $41.85\,W/m^2$. The percentage of each contribution is listed in Fig.\ref{fig:pie}. 
	Now we focus on the ice/concrete pad and consider only conduction. The steady-state conditions provide for the heat flow inside the slab
	\be\label{qpad}
	\dot{q}=\frac{T_S-T_p}{R_{tot}}=\frac{T_S-T_p}{R_{ice}+R_{conc}}\,,
	\ee
where $T_p$ is the temperature at the top of the pipes and the thermal resistances of the ice and concrete slabs are written as
\bea
&& R_{ice}=\frac{L}{k_{ice}}=1.33\times10^{-2}\,\frac{Km^2}{W}\,,\\
&& R_{conc}=\frac{d}{k_{conc}}=1.67\times10^{-2}\,\frac{Km^2}{W}\,,
\eea
where $k_{ice}=2.25\, W/mK$ and $k_{conc}=1.8\, W/mK$.
	The temperature at the top of the pipes $T_p$ is then estimated from
	\be
	\dot{q}=41.85\,W/m^2=\frac{T_S-T_p}{R_{tot}}\,,
	\ee
	which gives $T_{p}$=-5.9\textdegree C and can be cross-checked with heat balance inside the concrete slab only,
	\be
	\dot{q}=\frac{T_I-T_p}{R_{conc}}\,.
	\ee
This in fact returns $\dot{q}=41.92\,W/m^2\approx 41.85\,W/m^2$.

	Let us finally check the agreement between the thermal camera measurement and the result (\ref{icesurf}) for the ice surface temperature. The heat transfer via conduction inside the ice pad is written as
	\be
	\dot{q}_I=\frac{T_S-T_I}{R_{ice}}\,.
	\ee
Since this also must be equal to $\dot{q}$, namely $\dot{q}_I=\dot{q}=41.85$ W/m2 because of the steady state conditions, substituting and solving with respect to the temperature at the ice/concrete interface we get
\be
T_I=T_S-R_{ice}\dot{q}=-5.2\,\mathrm{^{\circ} C}\,,
\ee
that is indeed consistent with the measured value.

\section{Analytical temperature profile for the ice pad}\label{analytical}

\begin{figure}[t]
\centering

\begin{minipage}[t]{0.45\linewidth}
\centering
\includegraphics[width=\textwidth]{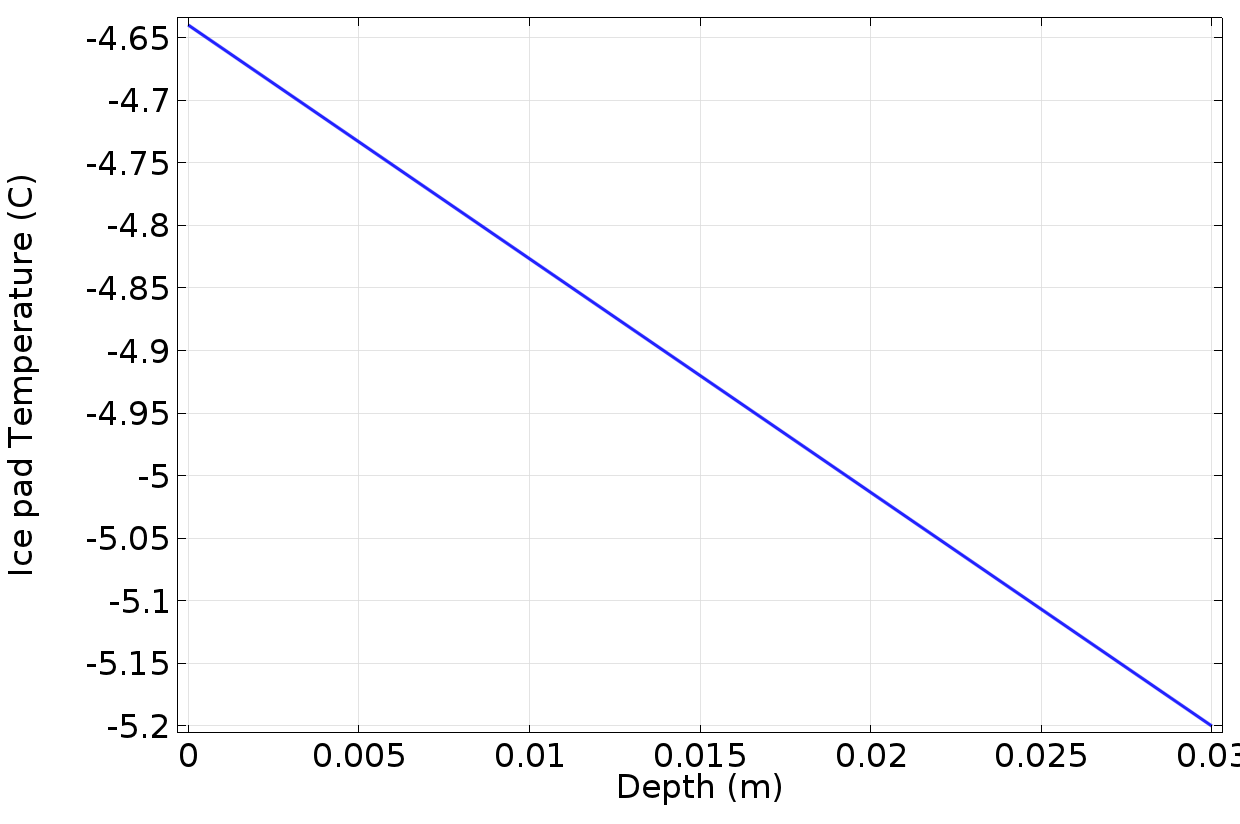}
		\caption{Initial condition $f(x)$.}
		\label{fig:f(x)}

\end{minipage}
\hspace{0.2cm}
\begin{minipage}[t]{0.45\linewidth}
\centering
\includegraphics[width=\textwidth]{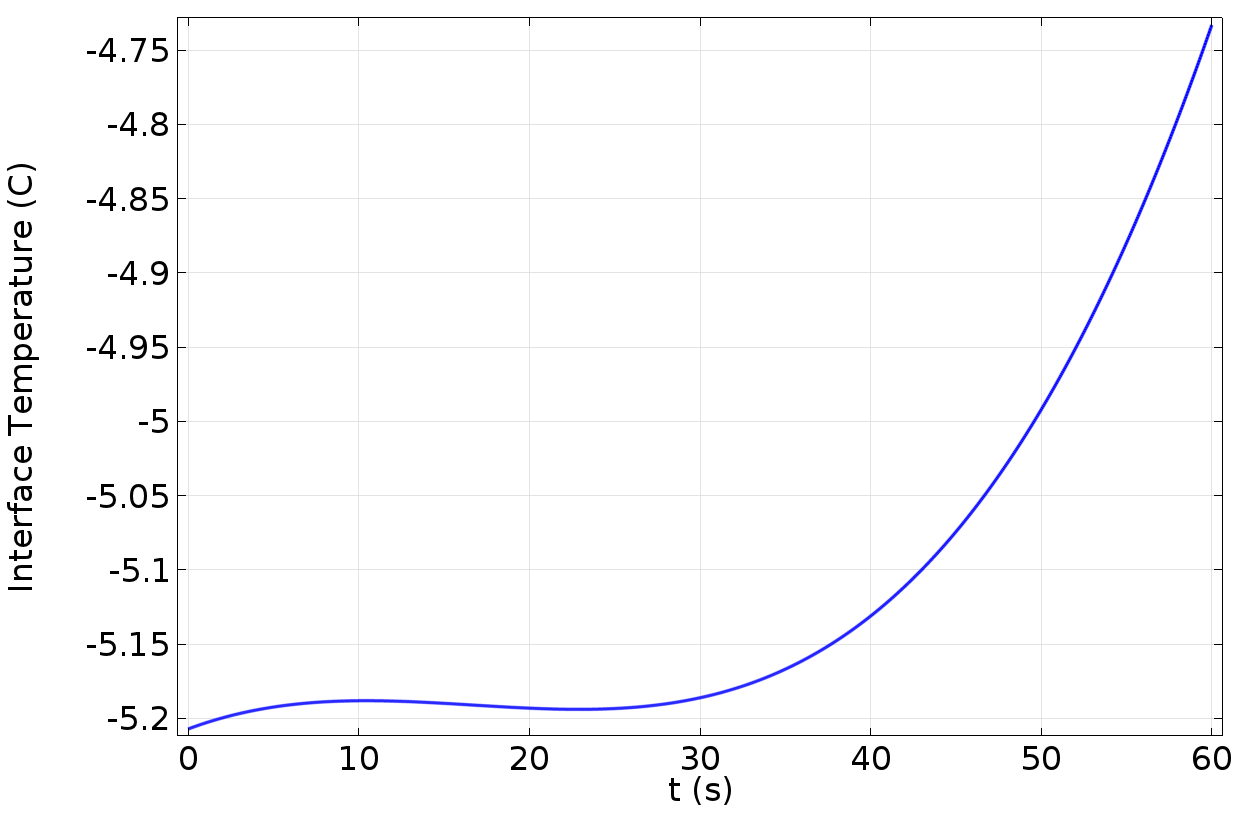}
		\caption{Interpolation of the interface temperature $T_I(t)$ for $t<60s$.}
		\label{fig:interpol3}
\end{minipage}

\end{figure}

In this section we derive an analytical formula for the temperature profile of the ice pad $T_{ice}(t,x)$, which can be used under any specific situation occurring in the ice hockey hall.
%
	The problem consists of solving the heat equation
	\be\label{heateqice}
	\frac{\partial u}{\partial t}=\alpha_I\frac{\partial^2 u}{\partial x^2}\,,
	\ee
	where $0<x<L=30mm$, and $\alpha_I$ is the thermal diffusivity of ice, with the time-dependent boundary conditions
	\bea
\label{bc1}&& u(0,t)=T_S(t)\,, \\
\label{bc2}&& u(L,t)=T_I(t)\,, 
\eea
and the initial condition
\be\label{ic}
u(x,t=0)=f(x)=18.67(0.03-x)-5.2\,, 
\ee
which is easily retrieved from the temperature data in the steady state regime. It gives indeed $T_S(0)$=-4.64\textdegree C at the ice surface and $T_I(0)$=-5.2\textdegree C at the ice/concrete interface, as in Fig.\ref{fig:f(x)}.

	The boundary conditions are in general given by the measurements. In this specific case, $T_S(t)$ is computed at the ice surface (at the water/ice interface) and $T_I(t)$ at the ice/concrete interface. They both are illustrated in Fig.\ref{fig:Tqzoom}.

\begin{figure}[t]
\centering
\begin{minipage}[b]{0.45\linewidth}
\centering
	\includegraphics[width=\textwidth]{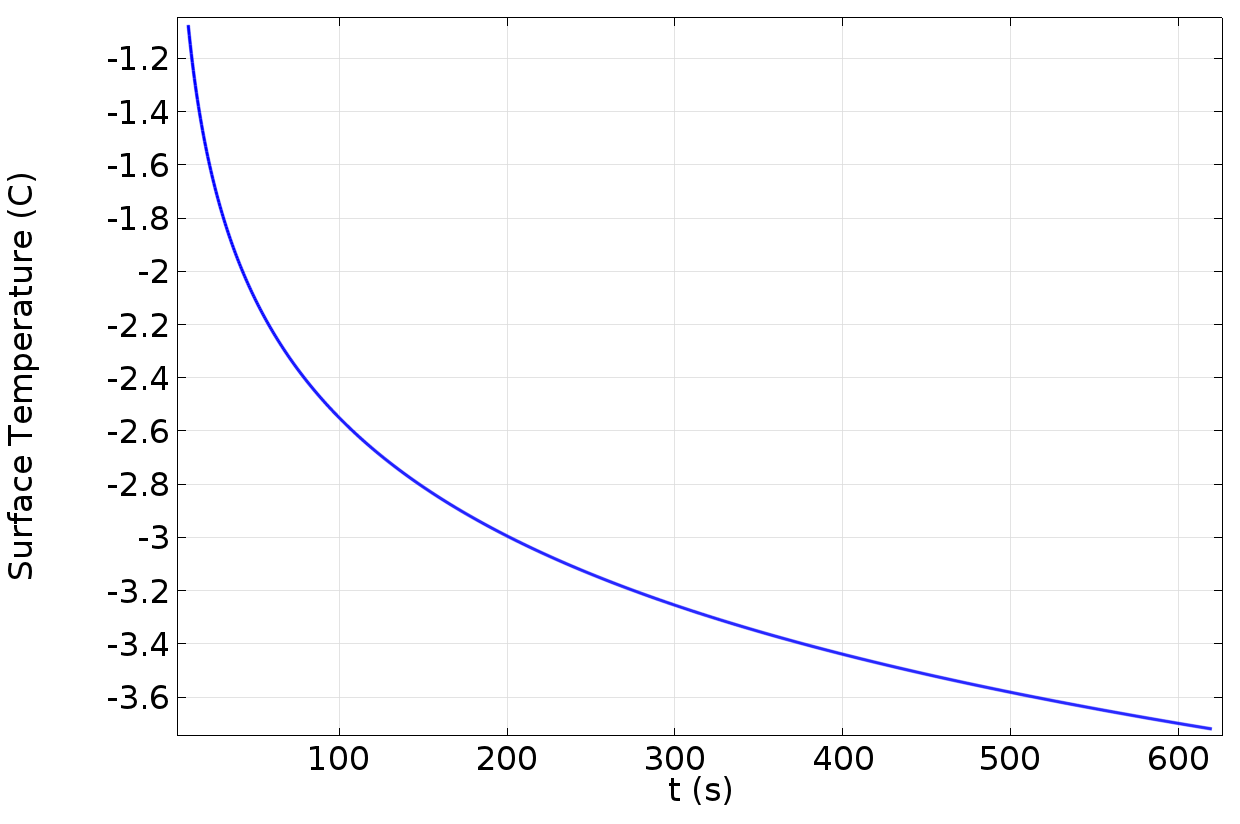}
		\caption{Interpolation of the surface temperature $T_S(t)$ over $t\leq620s$.}
		\label{fig:interpol1}

\end{minipage}
\hspace{0.1cm}
\begin{minipage}[b]{0.45\linewidth}
\centering
\includegraphics[width=\textwidth]{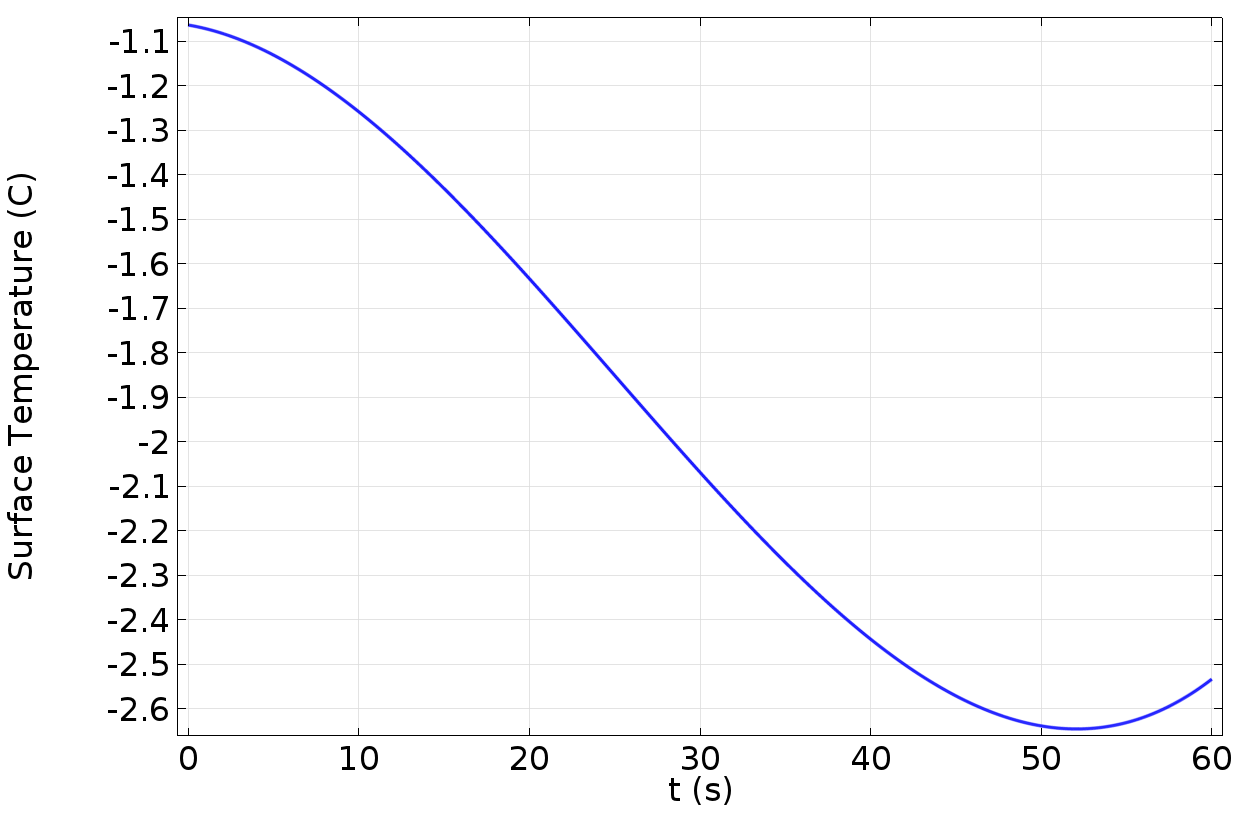}
		\caption{Interpolation of the surface temperature $T_S(t)$ for $t<60s$.}
		\label{fig:interpol2}

\end{minipage}

\end{figure}

	To obtain the analytical form of $T_S(t)$, we interpolate the temperature of the ice pad at the surface. The overall trend, including the entire curve from the beginning of resurfacing to $t=620s$, is clearly logarithmic. It gives the equation
\be
T_S(t)=-0.641\ln(t)+0.4016\qquad t[s]\leq 620\,.
\ee
If instead we consider e.g. only the first minute, the surface temperature becomes
\bea\label{TS}
T_S(t)=
2\times 10^{-5}t^3 - 0.0015t^2 - 0.0065t - 1.0633\,, \qquad
t[s]\leq 60\,.
\eea
These interpolations are given in Figs.\ref{fig:interpol1} and \ref{fig:interpol2}. The temperature raise in the last 10$s$ is due to the third order polynomial (this is non physical, still it remains within the experimental error and does not affect the result sensibly).
	The formula for the ice-concrete interface temperature, in the first approximation, is also a simple third-order polynomial obtained from the field measurements (Fig.\ref{fig:Tqzoom}),
\be\label{TI}
 T_I(t)= 6\times 10^{-6}t^3 - 3\times 10^{-4}t^2 + 4.3\times 10^{-3}t - 5.2071\,,
\ee
which is plotted in Fig.\ref{fig:interpol3}. We will apply these in the Appendix.
%
%

		To solve the Cauchy problem given by Eqs.(\ref{heateqice}), (\ref{bc1}), (\ref{bc2}) and (\ref{ic}), we adopt the method of Eigenfunctions Expansions detailed in \cite{TitchmarshI,TitchmarshII,Hahn} as follows. First impose the following \emph{Ansatz}
		\be
		w(x,t)=T_S(t)+x\left( \frac{T_I(t)-T_S(t)}{L} \right)\,,
		\ee
		that implies
		\bea
		&&w(0,t)=T_S(t)\,,\\
		&&w(L,t)=T_I(t)\,.
		\eea
Now define the temperature profile $u(x,t)$ as the sum
\be\label{decomp}
u(x,t)=w(x,t)+v(x,t)\,,
\ee
and substitute this expression in the Cauchy problem. We obtain
\be
v_t=\alpha_I v_{xx}-w_t\,,
\ee
where we have simplified the derivatives notation, with the Dirichlet boundary conditions $v(0,t)=v(L,t)=0$ and the initial condition $v(x,0)=f(x)-w(x,0)$. Now use the Eigenfunction expansion
\be
v(x,t)=\sum_{n=1}^\infty \hat{v}_n(t) \sin{(\lambda_nx)}\,,
\ee
to separate space and time dependence. The eigenvalues and eigenfunctions associated to the Dirichlet b.c. are
\bea
&&\lambda_n=\left( \frac{n\pi}{L} \right)\,, \;\; n\in\mathbb{N}\,; \qquad X_n(x)=\sin{(\lambda_nx)}\\
&& S(x,t)=-w_t=-(\dot{T}_I(t)-\dot{T}_S(t))\left(\frac{x}{L} \right)-\dot{T}_S(t)\,.
\eea
Therefore we obtain a first order linear ODE
\be\label{ODE}
\frac{d\hat{v}_n}{dt}+\alpha_I\lambda_n^2\hat{v}_n=\hat{S}_n(t)\,,
\ee
where
\bea\label{S}
\hat{S}_n(t)&=&\frac{2}{L}\int^L_0\left[ -\dot{T}_S(t)-\frac{x}{L}\left(\dot{T}_I(t) -\dot{T}_S(t)\right)\right]\sin\left(\frac{n\pi x}{L} \right)dx\nonumber\\
&=& \frac{2}{n\pi}\left[\left( \frac{\sin{n\pi}}{n\pi}-1 \right)\dot{T}_S(t)
-\left(\frac{\sin{n\pi}}{n\pi}-\cos{n\pi}
\right)
\dot{T}_I(t)  \right]\,,
\nonumber\\
\eea
with an integrating factor $F(t)=e^{\alpha_I\lambda_n^2t}$.
	Eq.(\ref{ODE}) can then be integrated to give the following solution,
\be
v(x,t)=\sum_{n=1}^\infty\left\{ \int_0^t d\tau e^{-\alpha_I\lambda_n^2(t-\tau)}\hat{S}_n(\tau)+e^{-\alpha_I\lambda_n^2t}c_n
\right\} \sin{(\lambda_nx)}\,,
\ee
where the coefficients
\be\label{cn}
c_n=\frac{2}{L}\int^L_0 \left\{ f(x)-\left[ (T_I(0)-T_S(0))\left( \frac{x}{L} \right) + T_S(0) \right] \right\} \sin\left(\frac{n\pi x}{L} \right)dx\,,
\ee
are given by the initial condition $u(0,x)=f(x)$.

	Putting everything together, we can now write the analytical temperature profile for the ice pad Eq.(\ref{decomp}) as
	\bea\label{Tprofile}
	u(x,t)&=&T_{ice}(x,t)=[T_I(t)-T_S(t)]\left( \frac{x}{L} \right)+T_S(t) \nonumber\\
	&+&\sum_{n=1}^\infty\left\{ \int_0^t d\tau e^{-\alpha_I\lambda_n^2(t-\tau)}\hat{S}_n(\tau)+e^{-\alpha_I\lambda_n^2t}c_n
\right\} \sin{(\lambda_nx)}\,,
	\eea
	that constitutes a novel result of this paper. This is a completely general formula, with implicit initial condition $f(x)$ and b.c. $T_I(t)$ and $T_S(t)$.
	
	The above clearly reduces to the ice temperature $T_S(t)$ at the surface for $x=0$, while it is slightly less immediate to verify that (\ref{Tprofile}) is consistent with the temperature at the bottom of the ice pad $T_I(t)$ for $x=L$. In this case we get
\bea
u(L,t)&=&T_{ice}(L,t)\equiv T_I(t)=[T_I(t)-T_S(t)]+T_S(t) \nonumber\\
	&+&\sum_{n=1}^\infty\left\{ \int_0^t d\tau e^{-\alpha_I\lambda_n^2(t-\tau)}\hat{S}_n(\tau)+e^{-\alpha_I\lambda_n^2t}c_n
\right\} \sin{(\lambda_nL)}\,.
\eea
Recall now that
	\be
	\lambda_n=\frac{n\pi}{L}\,,
	\ee
which implies
\be
-2T_S(t)\sum^\infty_1\frac{\sin{(\lambda_nL)}}{n\pi}=-2T_S(t)\sum^\infty_1\frac{\sin{n\pi}}{n\pi}=0\,,
\ee
since each term in the summation is identically zero. Regarding the term containing the integral,
\bea
2 \sum^\infty_1 \frac{\alpha_I\lambda_n^2}{n\pi}\sin{(\lambda_nL)}
	e^{-\alpha_I\lambda_n^2 t}\int^t_0 d\tau e^{\alpha_I\lambda_n^2 \tau} T_S(\tau)\propto
	\sum^\infty_1
n\sin{n\pi}e^{-\alpha_I\lambda_n^2 t}\equiv0\,,
\eea
again because $\sin{n\pi}=0,\;\forall n\in \mathcal{N}$. So we get an identity $0=0$, as required.

	Fisically, the terms in brackets in Eq.(\ref{Tprofile}) give the transient state correction to temperature, that depends on the history of the process (via the $t$-integral) and on the initial and boundary conditions by virtue of Eqs.(\ref{S}) and (\ref{cn}). This is plotted in Fig.\ref{fig:Tcorrection} for our specific example calculated in the Appendix, where the initial condition (\ref{ic}) and the boundary conditions (\ref{TS}) and (\ref{TI}) pertain the resurfacing process.
	
\begin{figure}[t]
\centering

\includegraphics[width=0.8\textwidth]{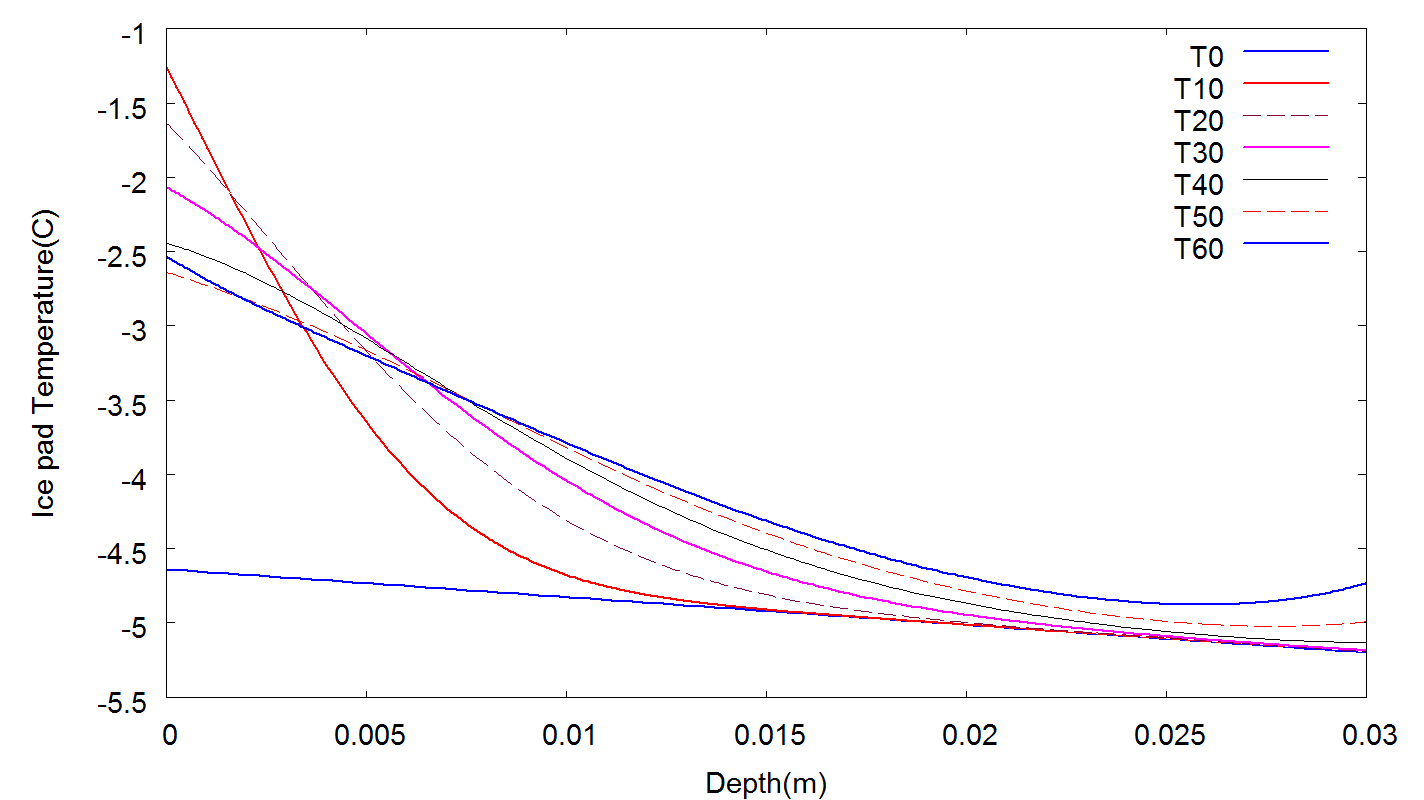}
\setlength{\abovecaptionskip}{3pt}
\caption{Temperature profile in the ice pad, analytical solution.}

\label{fig:analytical}

\includegraphics[width=0.75\textwidth]{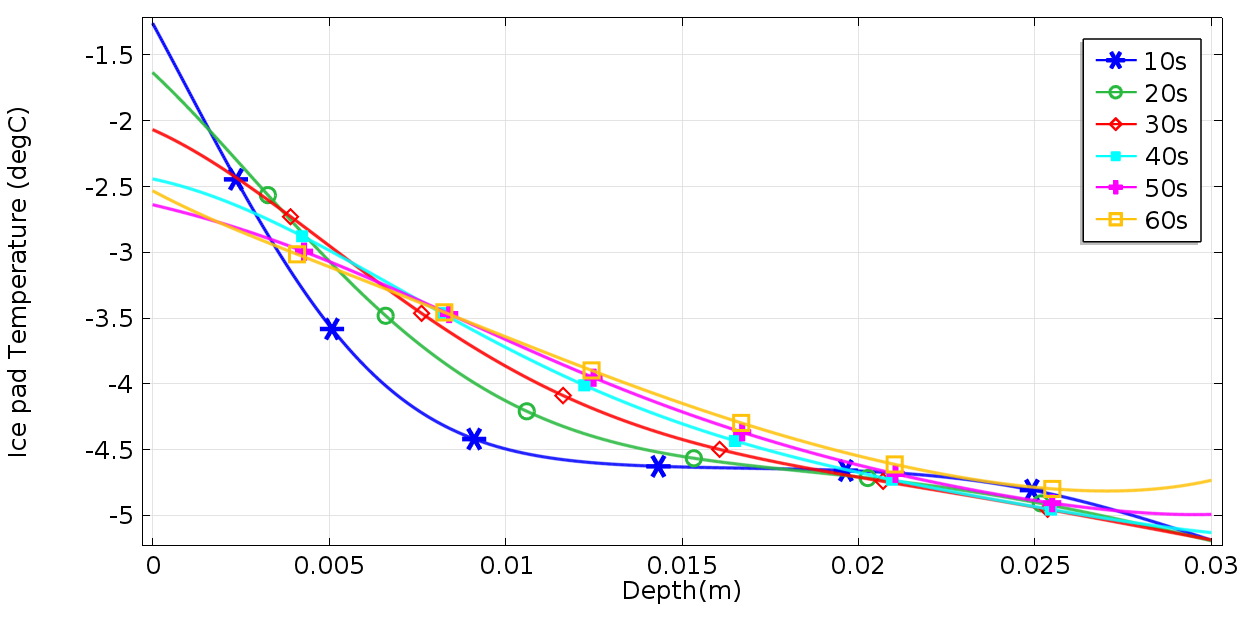}
\caption{FEM solution for $t=0...60s$.}
\label{fig:comsol}

\includegraphics[width=0.75\textwidth]{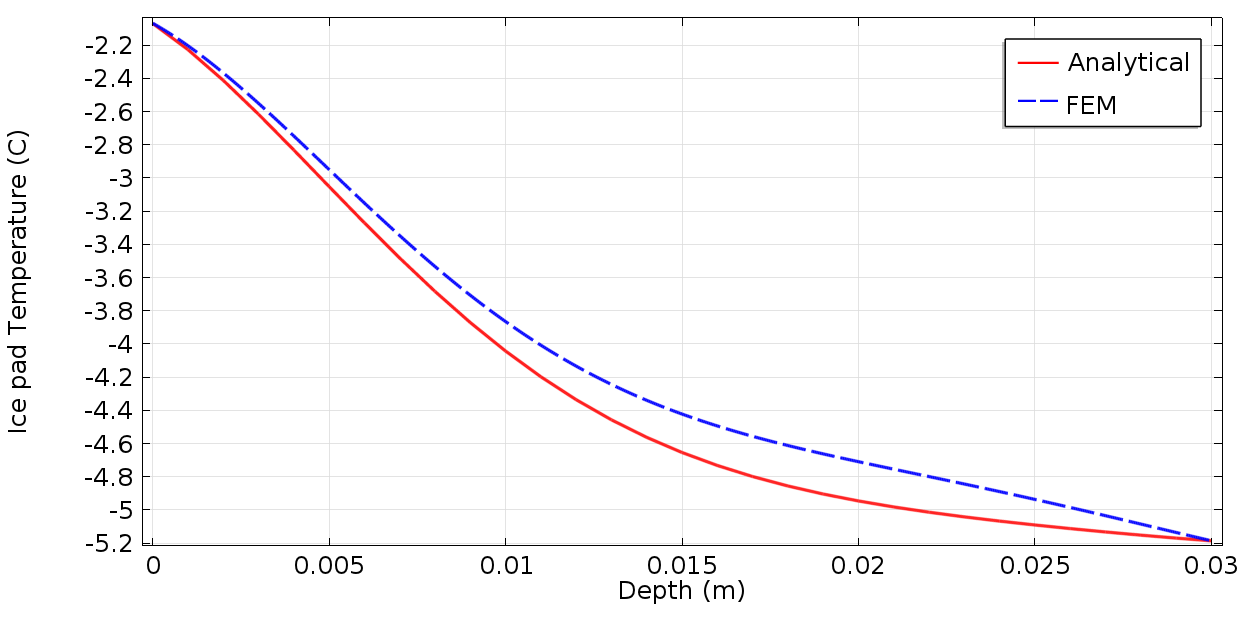}
\caption{Temperature profile after $t=30s$, analytical (solid) versus FEM (dashed).}
\label{fig:feman30}

\end{figure}

	The according analytical temperature profile (\ref{Tprofile}) is given in Figure \ref{fig:analytical}. In Fig.\ref{fig:feman30} this is compared to a numerical solution by a Finite Element Model (FEM), which is plotted in Figure \ref{fig:comsol}. Details of the analytical computation are discussed in the Appendix.
	%
	
	In conclusion, Eq.(\ref{Tprofile}) provides a general formula for the temperature profile $T(x,t)$ at any point $x$ of the ice pad, at any generic time $t$ corresponding to any situation in the ice hockey hall (closing hours, resurfacing or skaters' activity). The practical problem of studying the complex physical processes over and under the ice pad is here avoided by virtue of the boundary conditions (i.e. experimental data), which encode all the involved phenomenology in a very simple analytical form.

\section{Conclusions}

In this paper we have considered thermophysical processes in an ice hockey arena during standard operation hours. Detailed heat flux, air and ice temperature and relative humidity data are provided and discussed in a quantitative analysis with specific focus on the maintenance (resurfacing) phase.
The indoor air data provided constitute a stratification mapping at several heights from the ice rink, from 4cm to 8m above the ice, together with temperature and heat flux measurements at the ice/concrete interface.

Outside the occupation hours, we find a strong temperature stratification in the air above the ice rink, which is compromised once the skaters start their activity. The effect of the ventilation system, delivering air at $~25$\textdegree C at about 5m above the ice, is instead independent of the skaters, yet it is critically affecting the air layers temperature. As it is shown in Figure \ref{fig:AirT}, our data read warmer air at 5m than at 8m. Therefore one should control the ventilation system to avoid energy dissipation, and/or use waste energy techniques as in \cite{Lu20113360,Lu201491}.

An energy balance calculation shows the different contributions to the heat load on the ice rink, finding confirmation in the literature. The thermal radiation from the ceiling results to be the largest contribution ($74\%$), followed by lighting ($14\%$), with excellent agreement between our calculations and the temperature and heat flux measurements (this also constitutes a method for checking the data consistency). Together with the data measurements discussed in Section \ref{measurements}, such quantitative knowledge can aid energy saving efforts in a wide range of studies, since these processes occur in an average-sized ice hall under standard operating conditions.

Our measurements were then used to derive a direct analytical formula for the temperature inside the ice pad, viewed as a medium which is subject to conduction with time-dependent boundary conditions. When applied to the case at hand for validation, we showed that this is fairly consistent with a numerical computation for the ice pad temperature along its thickness, and that it reproduces the initial ice pad temperature profile correctly. Our formula is general and structurally simple, thus it can be readily applied to a number of investigations not limited to ice hockey halls.

%
	%
	Furthermore, we adopted a bottom-up approach where all the physics is encoded in the boundary conditions, circumventing an otherwise involved phenomenological analysis.
	This article suggests therefore a methodology which is firmly grounded on experimental data, using induction to obtain theoretical (predictive) results, and deduction to apply these formulas and check the data consistency.

	The accurate measurements, energy balance analysis and analytical temperature profile in the ice pad presented in this work can constitute useful tools for increasing the energy efficiency in the ice hockey arenas, since the ice thickness covers a role in the overall energy demand, as suggested by \cite{Somrani20081687}, and control of the indoor air stratification is capable of reducing the energy consumption appreciably, as demonstrated by \cite{en12040693}.



\section*{Acknowledgements}

This paper is dedicated to the memory of Professor Martti Viljanen. The authors acknowledge financial support by the Estonian Research Council with Institutional research funding grant IUT1-15 and by the Estonian Centre of Excellence in Zero Energy and Resource Efficient Smart Buildings and Districts, ZEBE, grant 2014-2020.4.01.15-0016 funded by the European Regional Development Fund.


\appendix
\section{Analytical temperature profile during resurfacing.}
The temperature profile (\ref{Tprofile}) was obtained in Section \ref{analytical} in implicit form. In this section we put it into context, by using the initial and boundary conditions Eqs.(\ref{ic}), (\ref{TS}) and (\ref{TI}) to obtain the temperature profile inside the ice pad during the resurfacing process. This constitutes also a validation of our formula, as the profile at $t=0$s agrees with the measurements reported in Section \ref{steadystate}.
First, we expand Eq.(\ref{S}),
\bea
\hat{S}_n(t)
&=& \frac{2}{n\pi}\left[\left( \frac{\sin{n\pi}}{n\pi}-1 \right)\dot{T}_S(t)
-\left(\frac{\sin{n\pi}}{n\pi}-\cos{n\pi}
\right)
\dot{T}_I(t)  \right]
\nonumber\\
&=&
\frac{1}{n\pi}\Bigg[
\left(8.4\frac{\sin{n\pi}}{n\pi}+3.6\cos{n\pi}-12
\right)\times 10^{-5} t^2\nonumber\\
&+&
\left(-0.0048\frac{\sin{n\pi}}{n\pi}-0.0012\cos{n\pi}+0.006
\right) t\nonumber\\
&+&
\left(-0.0261\frac{\sin{n\pi}}{n\pi}+0.0086\cos{n\pi}+0.013
\right)
\Bigg]\,,
\eea
then the coefficients (\ref{cn}),
\bea
c_n&=&\frac{2}{L}\int^L_0 \left\{ f(x)-\left[ (T_I(0)-T_S(0))\left( \frac{x}{L} \right) + T_S(0) \right] \right\} \sin{\left(\frac{n\pi x}{L} \right)}dx
\nonumber\\
&\sim& \frac{1}{n\pi}\left[7.167\left(\frac{\sin{n\pi}}{n\pi}-1\right)-0.0142\cos{n\pi}\right]\,,
\eea
because $0.998\sim1$. Already at this stage we notice the suppression factor $1/n$.
The integral in Eq.(\ref{Tprofile}) is now computed as
\bea\label{In}
&&I_n(t)\equiv \int_0^t d\tau e^{-\alpha_I\lambda_n^2(t-\tau)}\hat{S}_n(\tau)
=
\nonumber\\
&&=
\frac{1}{n^3}
\Bigg\{
(e^{-0.01327n^2t}-1)\Bigg[
\left(0.1649-\frac{2.762}{n^2} -\frac{7.284}{n^4}\right)\frac{\sin{n\pi}}{n}
\nonumber\\
&&-
\left(0.206+\frac{2.169}{n^2}+\frac{9.808}{n^4}\right)\cos{n\pi}
-0.312+\frac{10.846}{n^2}+\frac{32.693}{n^4}
\Bigg]
\nonumber\\
&&
+
\left(0.00064 \frac{\sin{n\pi}}{n} +0.00086\cos{n\pi}-0.0029\right)t^2
\nonumber\\
&&-
\Bigg[\left(0.037+\frac{0.097}{n^2}\right) \frac{\sin{n\pi}}{n} +\left(0.029+\frac{0.13}{n^2}\right)\cos{n\pi}-0.144-\frac{0.434}{n^2}
\Bigg]t
\Bigg\}
\nonumber\\
\eea
since $e^{\alpha_I\lambda_n^2\tau}=e^{0.01327n^2t}$.
\begin{figure}[t]
\centering

\begin{minipage}[b]{0.45\linewidth}

\centering
\includegraphics[width=\textwidth]{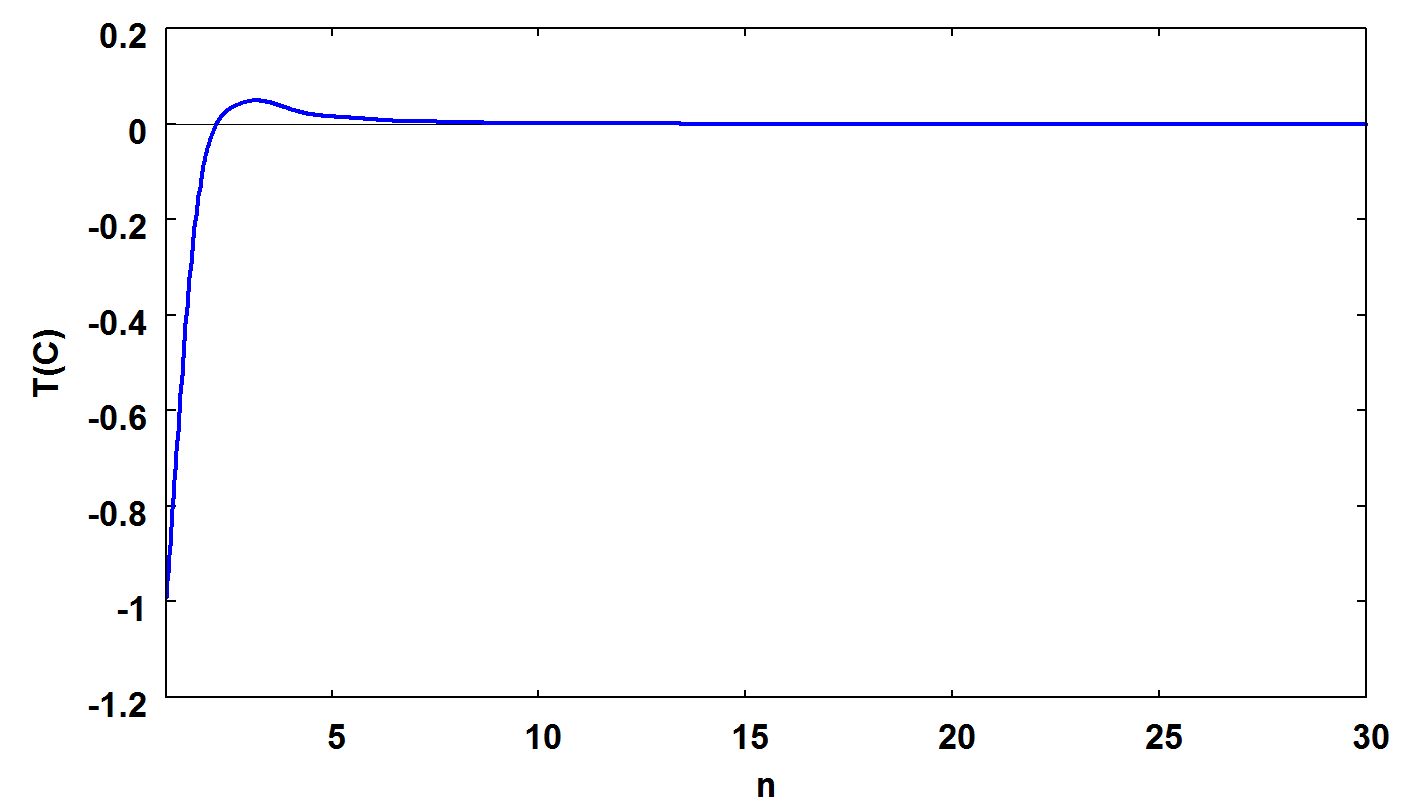}
\caption{Contribution of the first 30 terms in the summation (\ref{summation}), at $t=30s$.}
\label{fig:f(n)}
\end{minipage}
\hspace{0.2cm}
\begin{minipage}[b]{0.45\linewidth}
\centering
\includegraphics[width=\textwidth]{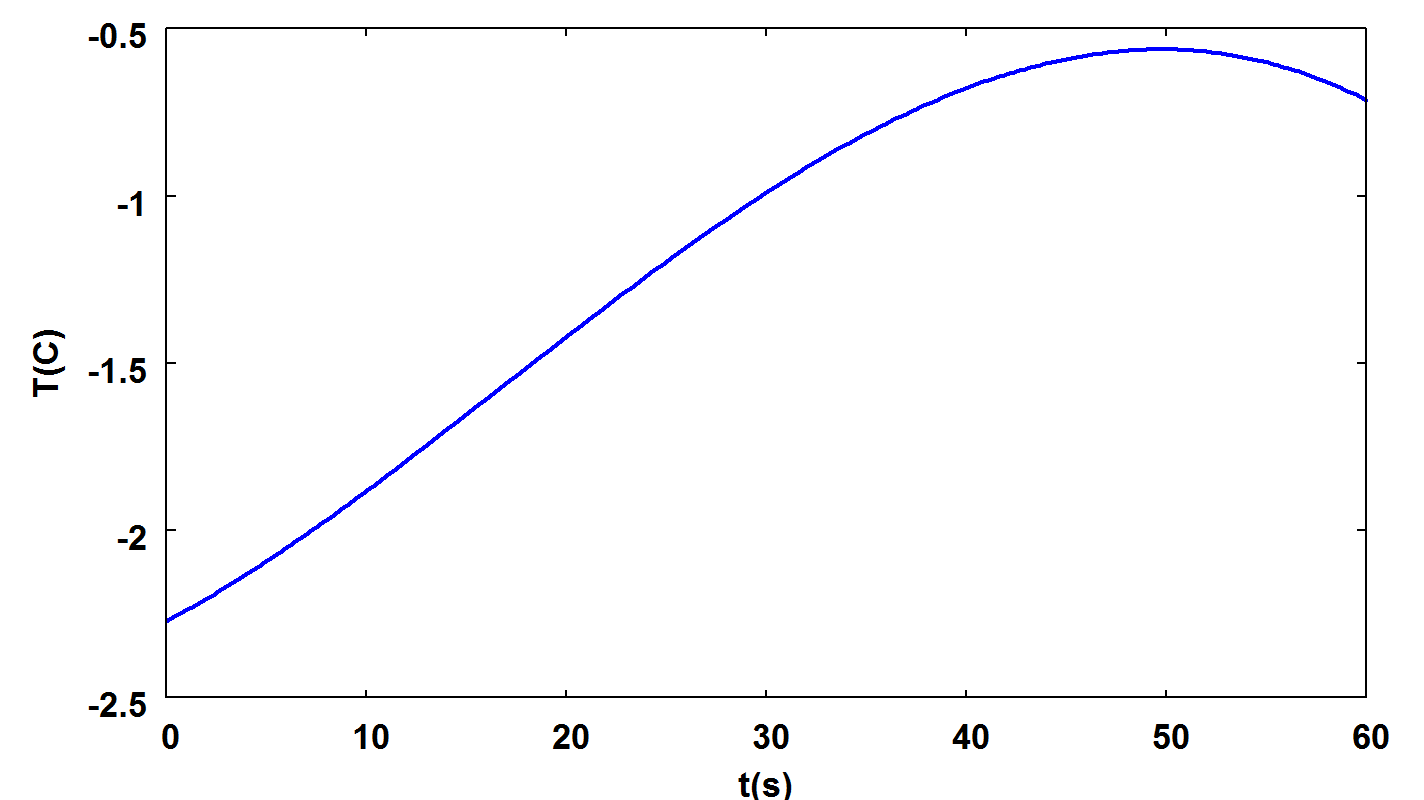}
\caption{Largest ($n=1$) contribution, where $0<t<60s$.}
\label{fig:f(t)}
\end{minipage}
\end{figure}
	We can thus recast the overall summation as
\bea\label{summation}
&&\sum_{n=1}^\infty\left\{ \int_0^t d\tau e^{-\alpha_I\lambda_n^2(t-\tau)}\hat{S}_n(\tau)+e^{-\alpha_I\lambda_n^2t}c_n
\right\} \sin{(\lambda_nx)}\nonumber
\\
&&=
\sum_{n=1}^\infty
 \left\{ I_n(t)+\dfrac{e^{-0.01327 n^2 t}}{n} \left(0.73\dfrac{\sin{n\pi}}{n}-0.01\cos{n\pi}-2.28\right)\right\}
\sin{\dfrac{n\pi}{L}x}\,,\nonumber
\\
\eea
which is clearly convergent, since everything is proportional to $\propto 1/n^a$.

The first 30 terms in the summation at $t=30s$ are plotted separately in Fig.\ref{fig:f(n)}; we see that only the first six or seven matter significantly. The largest contribution, for $n=1$, is shown in Fig.\ref{fig:f(t)}, where the correction to the temperature (in absolute value) is maximal at $t=0s$ and minimal at $t\sim 50s$.

\begin{figure}[t]
\centering
\includegraphics[width=0.8\textwidth]{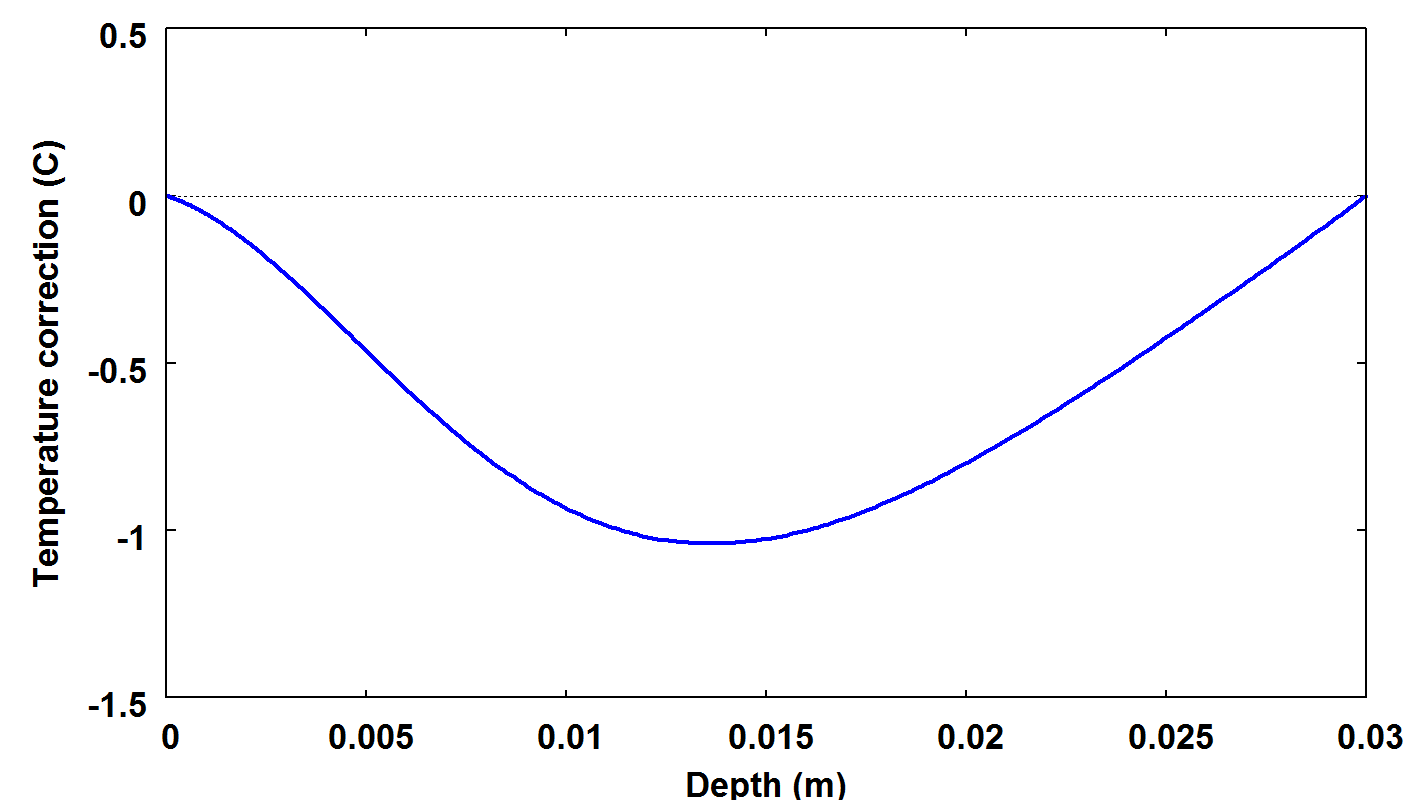}
\caption{Temperature correction at $t=30s$ with $n=1...100$ in Eq.(\ref{summation}).}
\label{fig:Tcorrection}
\end{figure}

	The temperature profile inside the ice pad Eq.(\ref{Tprofile}) is accordingly rewritten as follows,
\bea\label{Tprofileres}
u(x,t)&=&T_{ice}(x,t)=[T_I(t)-T_S(t)]\left( \frac{x}{L} \right)+T_S(t) \nonumber\\
	&+&\sum_{n=1}^\infty\left\{ \int_0^t d\tau e^{-\alpha_I\lambda_n^2(t-\tau)}\hat{S}_n(\tau)+e^{-\alpha_I\lambda_n^2t}c_n
\right\} \sin{(\lambda_nx)}\nonumber\\
&=&
-(1.4\times 10^{-5}t^3-0.0012t^2-0.0108t+4.1438)\frac{x}{L}
\nonumber\\
&&+2\times 10^{-5}t^3-0.0015t^2-0.0065t-1.0633
\nonumber\\
	&+&\sum_{n=1}^\infty\left\{ \int_0^t d\tau e^{-\alpha_I\lambda_n^2(t-\tau)}\hat{S}_n(\tau)+e^{-\alpha_I\lambda_n^2t}c_n
\right\} \sin{(\lambda_nx)},
\eea
where the temperature correction generated by the summation is computed with (\ref{In}) and (\ref{summation}); it is shown in Figure \ref{fig:Tcorrection} after 30$s$, where the first 100 terms are summed.
	Figure \ref{fig:feman30} compares the analytical solution (\ref{Tprofileres}) to an FEM calculation for $t=30s$, with $n=1...100$. We notice a good agreement between the two curves.

We remark that this result is specifically valid for any $t\leq60s$, since the boundary conditions $T_S(t)$ and $T_I(t)$ are interpolations corresponding only to such time interval. Choosing different times will change the explicit form of both $T_S(t)$ and $T_I(t)$. On the contrary, the temperature profile Eq.(\ref{Tprofile}) found in Section \ref{analytical} is given with implicit initial and boundary conditions, which makes it general and applicable to a range of diverse engineering problems.

\bibliography{Paper}
\bibliographystyle{natbib}

\end{document}